\documentclass{article}
\usepackage{a4wide}
\usepackage{epsfig}
\usepackage{natbib}

\newcounter{nref}
\newcommand{\bbib}{%
  \renewcommand{\refname}{\large\bf References}%
  \begin{thebibliography}{99}%
  \setcounter{enumiv}{\thenref}}
\newcommand{\ebib}{%
  \end{thebibliography}%
  \setcounter{nref}{\arabic{enumiv}}}
\newcommand{\head}[3]{%
  \setcounter{nref}{0}%
  \thispagestyle{empty}%
  \section*{\LARGE\bf #1}%
  \addcontentsline{toc}{section}{#1}%
  \large\itshape%
  #2\\\vspace{0.1pt}\\%
  #3%
  \normalsize\upshape%
  \bigskip}

\newcommand{\Msol}{\rm{M_{\odot}}}
\newcommand{\Ni}{\rm{^{56}Ni}}

\newcommand{\vnimax}{{v_{\rm Ni}^{\rm max}}}
\newcommand{\apj} {ApJ}
\newcommand{\apjl}{ApJ}
\newcommand{\aap} {A\&A}
\newcommand{\aaps} {A\&AS}
\newcommand{\mnras}{MNRAS}

\begin{document}


\head{The First Five Minutes of a Core Collapse Supernova:
      Multidimensional Hydrodynamic Models }
     {K.\,Kifonidis$^1$, T.\,Plewa$^{2,1}$, 
      H.-Th.\,Janka$^1$, E.\,M\"uller$^1$}
     {$^1$ Max-Planck-Institut f\"ur Astrophysik,
      Karl-Schwarzschild-Strasse 1, D-85740 Garching, Germany\\
      $^2$ Nicolaus Copernicus Astronomical Center, Bartycka 18, 
      00716 Warsaw, Poland}

\section{Introduction}

Numerous observations of SN 1987\,A suggest that extensive mixing has
taken place in the exploding envelope of the progenitor star
Sk~-69~202.  The early detection of X and $\gamma$-rays, the broad
profiles of infrared Fe and Co lines, as well as the shape of the
light curve cannot be explained without assuming that clumps of newly
synthesized $\Ni$, from layers close to the collapsed core,
have penetrated into the hydrogen envelope (see the reviews
of \citealt{ABKW89} and \citealt{Mueller98}, and the references therein).
That such mixing is probably generic in core collapse supernovae is
indicated by spectroscopic studies of SN 1987\,F, SN 1988\,A, SN
1993\,J \citep[and references therein]{Spyromilio94} and SN 1995\,V
\citep{F+98}.  Furthermore, it might also explain the detection of
fast moving clumps of metal-enriched material in the
Vela~\citep{AET95}, Cas~A~\citep{A+94} and Puppis~A \citep{WK85}
supernova remnants as well as the isotopic composition of specific SiC
grains with possible supernova origin found in primitive meteorites
(\citealt{NAZWL96}; Amari this volume).  But even more important from
the point of view of supernova modellers is the fact, that a detailed
understanding of the problem of nucleosynthesis and mixing can give us
invaluable information about the explosion mechanism itself.  Thus,
the observations have instigated theoretical work on multidimensional
supernova models which focused either on the role of convection
occuring within the first second of a delayed, neutrino-driven
explosion \citep{Mezz+98,JM96,BHF95,HBFC94,Mil93}, or on the growth of
Rayleigh-Taylor instabilitites during the late evolutionary stages
\citep{NSS98,HB92,MFA91,YS91,HMNS90}.

However, multidimensional simulations which follow the evolution of
the stalled supernova shock from its revival due to neutrino heating,
until its emergence from the stellar surface have not yet been
performed.  Due to the presence of vastly different spatial and
temporal scales and the range of physical processes involved during
the early stages following core collapse, all studies of mixing in
core collapse supernovae have hitherto neglected the influence of
neutrino-driven convection in seeding the Rayleigh-Taylor
instabilities.  Instead, a shock wave was created artificially by
depositing the explosion energy near the center of a pre-collapse
progenitor model and following the propagation of the shock in one
spatial dimension until it had reached one of the unstable composition
interfaces. This was either chosen to be the He/H interface in case of
Type~II supernova models (e.g. \citealt{HB92}; \citealt{MFA91}) or the
C+O/He interface in case of Type~Ib models \citep{HMNS94}.  Only then
were the 1D models mapped to a 2D grid and the rest of the evolution
followed with a multidimensional code.  A somewhat different,
two-dimensional approach, was chosen in the recent calculations of
\cite{NHSY98} who initiated the explosion using a parameterized,
aspherical shock wave and computed the resulting nucleosynthesis using
a marker particle approximation. These models have been subsequently
used by \cite{NSS98} for a study of Rayleigh-Taylor instabilities at
the He/H interface.  Still, however, these calculations do not address
the complications introduced by the explosion mechanism and thus
suffer from a number of assumptions.  Furthermore, their numerical
resolution appears to be hardly sufficient to resolve instabilities
which occur within the first minutes of the evolution.

In the present contribution a first step towards a more consistent
multidimensional picture of core collapse supernovae is attempted
by trying to answer the questions
\begin{itemize}
\item
 What happens in the first minutes and hours of a core collapse
 supernova {\em if\/} neutrino heating
 indeed succeeds in reviving the supernova shock?
\item
 Can neutrino-driven convection in conjunction with the later
 Rayleigh-Taylor instabilities lead to the high iron velocities 
 observed in SN~1987\,A?
\end{itemize}

For this purpose we have carried out high-resolution 2D supernova
simulations which for the first time cover the neutrino-driven
initiation of the explosion, the accompanying convection and
nucleosynthesis as well as the Rayleigh-Taylor mixing. In the
following, we present preliminary results from these
calculations, focusing on the first $\sim 300$ seconds of
evolution. A summary of our work can also be found in \cite{KPJM99}.

\section{Numerical Method and Initial Data}
\label{sect:numerics}

We split our simulation into two stages.  The early evolution ($t
\leq 1$\,s) which encompasses shock revival by neutrino heating,
neutrino-driven convection and explosive nucleosynthesis is followed
with a version of the {\sc HERAKLES} code (T. Plewa \& E. M\"uller, in
preparation).  This hydrodynamics code solves the multidimensional
hydrodynamic equations using the direct Eulerian version of the
Piecewise Parabolic Method \citep{CW84} augmented by the Consistent
Multifluid Advection (CMA) scheme of \cite{CMA} in order to guarantee
exact conservation of nuclear species.  We have added the input
physics (neutrino source terms, equation of state, boundary
conditions, gravitational solver) described in \cite{JM96} (henceforth
JM96) with the following modifications.  General relativistic
corrections are made to the gravitational potential following
\cite{Van_Riper}.  A 14-isotope network is incorporated in order to
compute the explosive nucleosynthesis.  It includes the 13
$\alpha$-nuclei from $\rm ^4He$ to $\Ni$ and a representative tracer
nucleus which is used to monitor the distribution of the
neutrino-heated, neutron-rich material and to replace the $\Ni$
production when $Y_{\rm e}$ drops below $\sim 0.49$
\citep[cf.][]{TNH96}.

We start our calculations 20\,ms after core bounce from a
model of \cite{Bruenn} who has followed core-collapse and bounce in
the 15\,$\Msol$ progenitor model of \cite{WPE88}. The model is mapped
to a 2D grid consisting of 400 radial zones ($ 3.17\times 10^{6}\,{\rm
cm} \leq r \leq 1.7\times10^9$\,{\rm cm}), and 180 angular zones ($0
\leq \theta \leq \pi$; cf. JM96 for details).  A random initial seed
perturbation is added to the velocity field with a modulus of
$10^{-3}$ of the (radial) velocity of Bruenn's post-collapse model.
The calculations are carried up to 885\,ms.  At this time the
explosion energy has saturated and essentially all nuclear reactions 
have frozen out.  We will henceforth refer to this calculation as our 
``explosion model''.

The subsequent shock propagation through the stellar envelope and the
growth of Rayleigh-Taylor instabilities is followed with the AMRA
Adaptive Mesh Refinement (AMR) code (T. Plewa \& E. M\"uller, in
preparation). This code uses a different variant of HERAKLES as its
hydrodynamics solver which does not include the neutrino physics.
Gravity is neglected in the AMR calculations since, as is the case for
the neutrino source terms, it does not influence the propagation of
the shock during late evolutionary stages.  However, gravity is
important for determining the amount of fallback, a problem which is
outside the scope of the present study.  The equation of state takes
into account contributions from photons, non-degenerate electrons,
$\rm e^+e^-$-pairs, $\rm ^1H$, and the nuclei included in the reaction
network. The AMR calculations are started with the inner and outer
boundaries located at $r_{\rm in}=10^8$\,cm (i.e. inside the hot
bubble containing the neutrino-driven wind) and $r_{\rm
out}=2\times10^{10}$\,cm, respectively.  No further seed perturbations
are added.  We use up to four levels of mesh refinement and refinement
factors of four in each grid direction, yielding a maximum resolution
equivalent to that of a uniform grid of $3072 \times 768$ zones.  In
order to keep the radial resolution as high as possible during any
given evolutionary time, we do not include the entire star but allow
the code to expand the radial extent of the base grid by a factor of 2
to 4 whenever the supernova shock is approaching the outer grid
boundary. The latter is moved from its initial value out to $r_{\rm
out}=1.1\times10^{12}$\,cm at $t=300$\,s.  Reflecting boundary
conditions are used at $\theta = 0$ and $\theta = \pi$ and free
outflow is allowed across the inner and outer radial boundaries.

\section{Nucleosynthesis and Neutrino-Driven Convection}

\begin{figure}[t]
\begin{center}
\epsfig{file=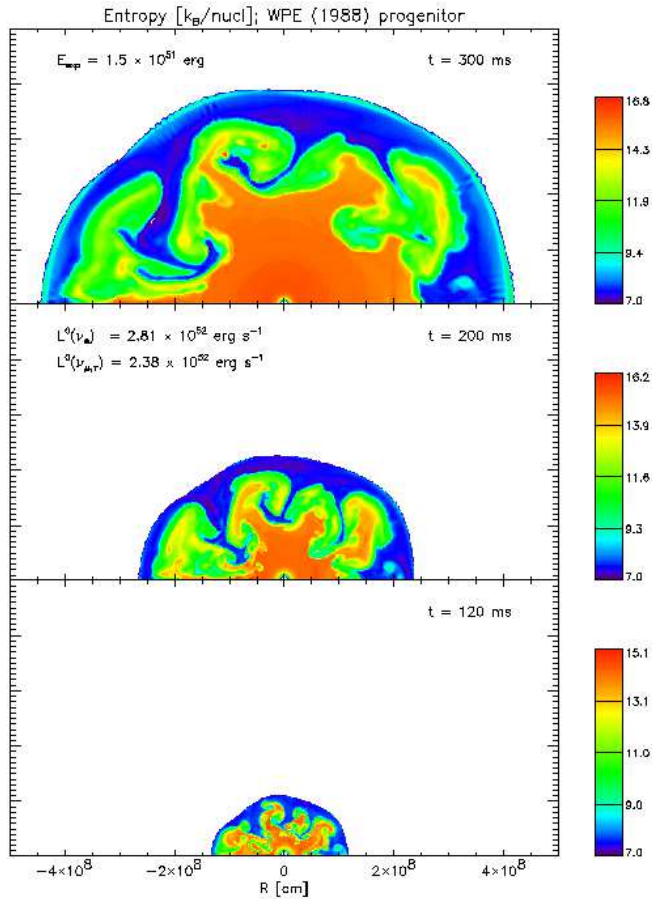,width=0.6\textwidth}
\end{center}
\caption{Evolution of the entropy in our explosion model within 
         the first 300\,ms after the start of the calculations.
         Neutrino-heated, deleptonized material
         rises in large bubbles and distorts the shock 
         (outermost discontinuity).}
\label{fig:entropy}
\end{figure}

The general features of our explosion model are comparable to the
models of JM96. For the initial neutrino luminosities, which are
prescribed at the inner boundary, somewhat below the neutrino sphere,
and decay with time as described in JM96, we have adopted a value of
$L_{\nu_e}^0 = 2.8125 \times 10^{52}\,{\rm erg/s}$ for the electron
neutrinos and $L_{\nu_x}^0 = 2.375\times10^{52}\,{\rm erg/s}$ for the
heavy lepton neutrinos (with $\nu_x = \nu_{\mu}, \bar \nu_{\mu},
\nu_{\tau}, \bar \nu_{\tau}$).  The parameters describing lepton and
energy loss of the inner iron core were set to $\Delta Y_l = 0.0875, \Delta
\varepsilon = 0.0625$ (cf. JM96). The neutrino spectra are the same as
in JM96.

\begin{figure}[t]
\begin{center}
\epsfig{file=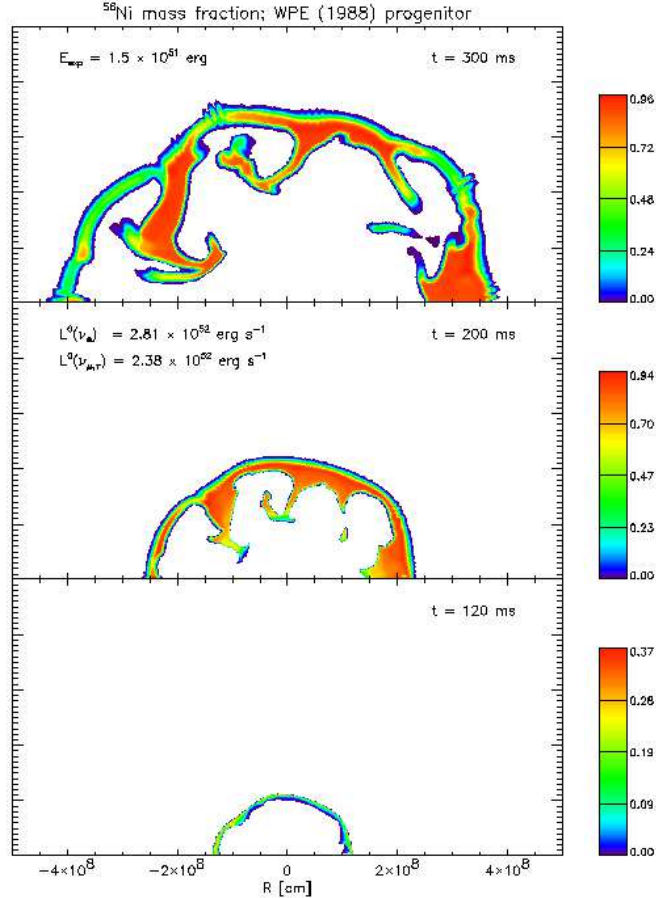,width=0.6\textwidth}
\end{center}
\caption{Evolution of the $\Ni$ mass fraction within 
         the first 300\,ms after the start of the calculations.
         $\Ni$ has only been synthesized in those regions which have an
         electron fraction $Y_e > 0.49$ and reached peak-temperatures 
         above $\sim 5\times10^9$\,K.
         }
\label{fig:nickel}
\end{figure}

For the chosen neutrino luminosities the shock starts to move out of
the iron core almost immediately.  Convection between shock and gain
radius sets in $\sim 30$\,ms after the start of the simulation in form
of rising blobs of neutrino-heated, deleptonized material (with
$Y_{\rm e} \ll 0.5$) separated by narrow downflows with $Y_{\rm
e} \approx 0.49$ (Fig.~{\ref{fig:entropy}}).
The shock reaches the Fe/Si interface at $r = 1.4\times10^8$\,cm after
$\sim 100$\,ms. Shortly thereafter, at $t\approx 120$\,ms temperatures
right behind the shock have dropped below $\approx 7 \times 10^9$\,K,
and $\Ni$ starts to form in a narrow shell (Fig.~\ref{fig:nickel}).
During the ongoing expansion and cooling, $\Ni$ is also synthesized in
the convective region. However, its synthesis proceeds exclusively in
the narrow downflows which separate the rising bubbles and have a
sufficiently high electron fraction $Y_{\rm e}$. This leads to a
highly inhomogeneous nickel distribution (middle panel of
Fig.~\ref{fig:nickel}) shortly before complete silicon burning freezes
out at $t \approx 250$\,ms.  Moreover, convective motions are still
present even after $\Ni$ production ceases, and distort the nickel
containing shell (upper panel of Fig.~\ref{fig:nickel}).  Only when
convection stops around $t \approx 400$\,ms, the flow pattern becomes
frozen in and the entire post-shock region expands nearly uniformly.

Due to the asphericity of the shock caused by the rising bubbles,
``bent'' shells containing the products of incomplete silicon burning
as well as oxygen burning form outside the nickel-enriched region.
The post-shock temperature drops below $2.8 \times 10^9$\,K at $t =
495$\,ms, when the shock is about to cross the Si/O interface. Thus,
our model shows only moderate oxygen burning (due to a non-vanishing
oxygen abundance in the silicon shell), and negligible neon and carbon
burning.  This is caused by the specific structure of the progenitor
model of \cite{WPE88} and may change when different (especially more
massive) progenitors are used. In total, $0.052\,\Msol$ of $\Ni$ are
produced, while $0.10\,\Msol$ of material in the deleptonized bubbles
are synthesized at conditions with $Y_{\rm e} < 0.49$ and end up as
neutron-rich nuclei. The explosion energy of our 2D model saturates at
$1.48\times10^{51}$\,erg at $t = 885$\,ms. This value is still to be
corrected for the binding energy of the outer envelope.

\section{Growth of Rayleigh-Taylor Instabilities}

During the next seconds the shock detaches from the formerly
convective shell that carries the products of explosive
nucleosynthesis, looses its asphericity and crosses the
C+O/He-interface.  Along with the slow-down which the shock
experiences after passing this interface due to the varying density
gradient, the entire post-shock material is also rapidly
decelerated. Twenty seconds after core-bounce, the metal-containing
shell has thus been compressed to a thin, dense layer which is bounded
inwards by a reverse shock and contains two regions which show crossed
density and pressure gradients.  The first of these is located at the
Ni+Si/O-interface while the second coincides with the
C+O/He-interface. Thus, Rayleigh-Taylor instabilities at the Ni+Si/O
and C+O/He-interfaces grow rapidly.

\begin{figure}[t]
\begin{center}
\epsfig{file=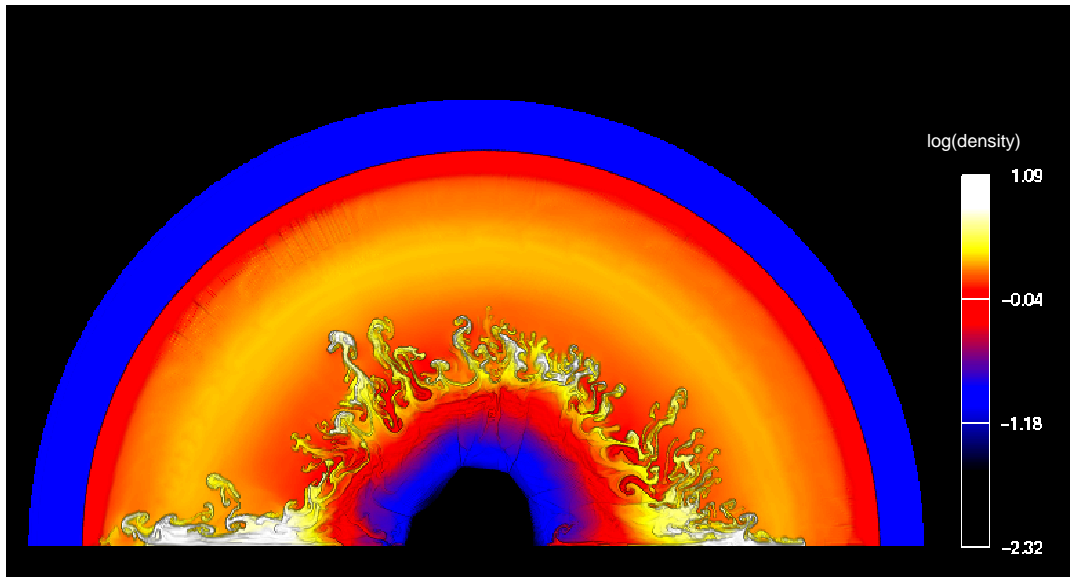,width=0.95\textwidth,
        bbllx=15,bblly=15,bburx=578,bbury=319}
\end{center}
\caption{Density distribution ($\log_{10} \rho~[{\rm g\,cm^{-3}}]$) in
         the exploding star, 100\,s  after core bounce.  The outer edge
         of the displayed domain (blue) coincides with a radius of 
         $1.15\times10^{11}$\,cm.
         The supernova shock (bright orange discontinuity near the outer
         edge) has crossed the He/H-interface and reached a radius of 
         $r = 10^{11}$\,cm. Rayleigh-Taylor mushrooms distribute the 
         products of explosive nucleosynthesis throughout the He-core.
         The reverse shock (aspherical discontinuity) 
         visible near the center was created after the main shock had 
         passed the C+O/He-interface of the star. 
         (The excessive outflows along the axis are specific to 
         the use of polar coordinates.)}
\label{fig:amra100}
\end{figure}

\begin{figure}[t]
\begin{center}
\epsfig{file=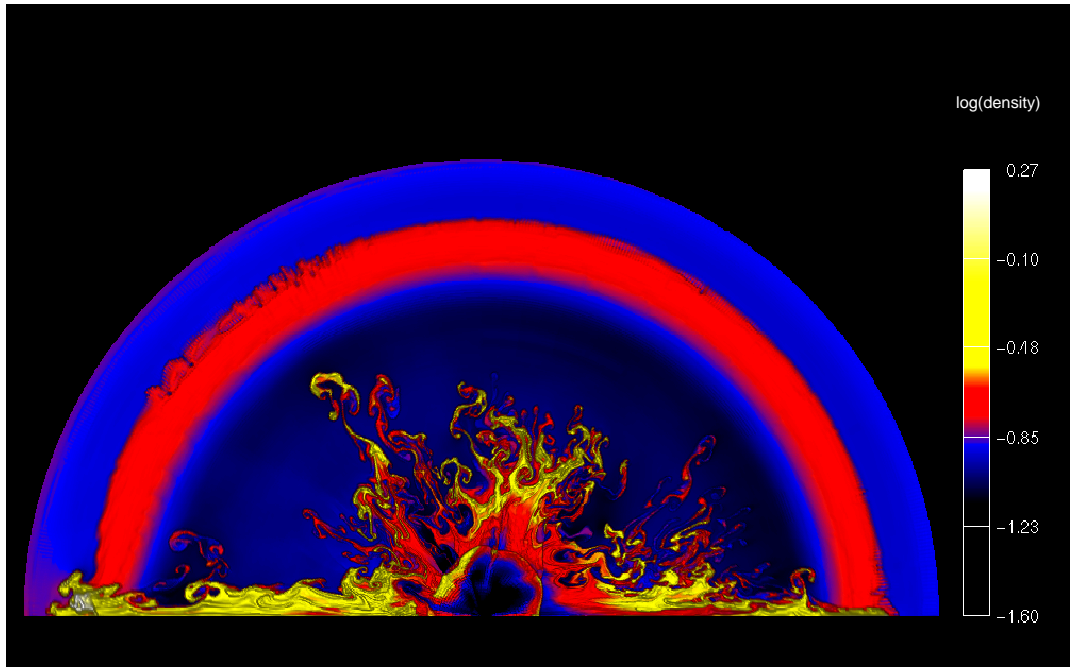,width=0.95\textwidth,
        bbllx=15,bblly=15,bburx=578,bbury=366}
\end{center}
\caption{Density distribution ($\log_{10} \rho~[{\rm g\,cm^{-3}}]$) 300\,s  after
         core bounce in the inner $\sim 3\times10^{11}$\,cm of the star.
         The supernova shock (outermost blue discontinuity) is located inside the
         hydrogen envelope at $r = 2.7\times10^{11}$\,cm. 
         A dense shell (visible as a bright red ring) 
         has formed behind the shock. Its outer boundary coincides with 
         the He/H interface while its inner boundary is in the process
         of steepening into a reverse shock. The fastest mushrooms
         have almost caught up with the inner boundary of the shell.
         (The excessive outflows along the axis are specific 
         to the use of polar coordinates.)}
\label{fig:amra300}
\end{figure}

At $t=100$\,s (Fig.~\ref{fig:amra100}) the instabilities are fully
developed and have already interacted with each other. Nickel and
silicon are dragged upward into the helium shell in rising mushrooms
on angular scales from $1^{\circ}$ to about $5^{\circ}$, whereas
helium is mixed inward in bubbles. Oxygen and carbon, located in
intermediate layers of the progenitor, are swept outward as well as
inward in rising and sinking flows. At $t = 300$\,s the densities
between the dense mushrooms and the ambient medium differ by factors
up to 5 while the fastest mushrooms have already propagated out to
more than half the radius of the He core (Fig.~\ref{fig:amra300}).

We observe that the nickel is born with very high velocities, of the
order of 15,000 km/s at $t=300$\,ms. These velocities drop
significantly during the subsequent evolution, especially after the
shock passes the C+O/He interface.  At a time of $t=50$\,s most of the
$\Ni$ is expanding with 3200 -- 4500\,km/s and maximum velocities
$\vnimax$ are around 5800\,km/s. During the following 50 seconds,
$\vnimax$ drops to $\sim 5000$~km/s.  However, after $t=100$\,s, the
clumps start to move essentially ballistically through the helium core
and only a slight decrease of $\vnimax$ to $\sim 4700$\,km/s occurs
until $300$\,s, when the bulk of $\Ni$ has velocities below
$3000$\,km/s.

We have recently accomplished to follow the subsequent evolution of
our model up to 16\,000\,s after core-bounce.  The unsteady
propagation speed of the supernova shock which has formerly induced
the instability at the Ni+Si/O and C+O/He-interfaces also leads to the
formation of a dense (Rayleigh-Taylor unstable) shell at the He/H
interface. As before, the inner boundary of this shell steepens into a
reverse shock (Fig.~\ref{fig:amra300}), while in the process of this
the entire shell is rapidly slowed down.  Our high-resolution
simulations reveal a potentially severe problem for the mixing of
heavy elements into the hydrogen envelope of Type II supernovae like
SN~1987A. We find that the fast nickel containing clumps, after having
penetrated through this reverse shock, dissipate a large fraction of
their kinetic energy in bow shocks created by their supersonic motion
through the shell medium. This leads to their deceleration to $\sim
2000$~km/s in our calculations. Contrary to all previous studies,
which tried to {\em accelerate\/} the nickel by the instabilities, our
computations show, that the main problem for obtaining high nickel
velocities is how to avoid {\em decelerating\/} the clumps once they
reach the He/H interface. The growth time scale of the instability at
this interface is too large to allow for a fast shredding of the dense
shell and the formation of ``holes'' through which the clumps could
penetrate more easily. The high iron velocities in SN 1987\,A might
therefore hint towards a smoother density profile exterior to the
He-core, which would suppress dense shell formation, or towards a
global asymmetry of the explosion.

During the computations we became aware of oscillations with angle in
parts of the postshock flow (Figs.~\ref{fig:nickel} and
\ref{fig:amra300}). These are caused by the ``odd-even decoupling''
phenomenon associated with grid-aligned shocks (\citealt{JJQ}; see also
\citealt{LeVeque98}). As a consequence, the maximum nickel velocities,
$\vnimax$, obtained in our AMR calculations have probably been
overestimated by $\sim 25\%$ because the growth of some of the
mushrooms was influenced by the perturbations induced by this
numerical defect. The main results of our study, however, are not
affected.

\section{Conclusions}
\label{sect:conlusions}

We have studied the evolution of a core collapse supernova in a
15\,$\Msol$ blue supergiant progenitor focusing on the first 300\,s
after core bounce. As a result of the interplay between
neutrino-matter interactions, hydrodynamic instabilities and
nucleosynthesis in the framework of the neutrino-driven explosion
mechanism, the products of explosive silicon and oxygen burning are
mixed throughout the inner half of the helium core of the progenitor
star. Ballistically moving, metal-enriched clumps with velocities up
to more than $\sim 4000$\,km/s are observed.  Rayleigh-Taylor
instabilities at the C+O/He and Ni+Si/O-interfaces transport helium
deep inward and sweep nickel, silicon, oxygen and carbon outward in
rising mushrooms.

Our simulations suggest that high-velocity metal-rich clumps are
ejected during the explosion of Type Ib (and Ic) supernovae. In case
of Type II supernovae, however, the dense shell left behind by the
shock passing the boundary between helium core and hydrogen envelope,
causes a substantial deceleration of the clumps.  Thus, we cannot
confirm the results of \cite{HB92} that ``premixing'' of the $\Ni$
within the first minutes of the explosion, and its later interaction
with the instability at the He/H-interface, can explain the high iron
velocities observed in SN~1987\,A. Moreover, our calculations strongly
indicate, that all computations of Rayleigh-Taylor mixing in Type~II
supernovae carried out so far (including the case of SN~1987\,A) have
been started from {\em overly simplified\/} initial conditions since
they have neglected clump formation during the first minutes of the
explosion. In future work, it will be most interesting to study how
the discussed effects depend on different progenitor models and to
explore the implications of the strong outward mixing of $\Ni$ for the
light curves and spectra of Type~Ib supernovae.

\section*{Acknowledgements}

We are very grateful to Stanford Woosley for providing us with 
the progenitor model used in this study as well as for stimulating 
discussions about Type~Ib supernovae.
We thank S. Bruenn for making available the data of 
his post-bounce model and P. Cieciel\,ag and R. Walder for their
help regarding visualization. The work of TP was partly supported by     
grant KBN 2.P03D.004.13 from the Polish Committee for Scientific
Research. The simulations were performed on the NEC SX-4B and 
CRAY J916/16512 of the Rechenzentrum Garching.


\end{document}